\documentclass{iaus}
\usepackage{graphicx}
\usepackage{color}
\usepackage{ulem}

\newcommand{\msun}{${\rm M}_{\odot}$}

\title[Galactic consequences of clustered star formation] 
{Galactic consequences of clustered star formation}

\author[M.R. Haas \& P. Anders]   
{M.R. Haas$^1$
 \and P. Anders$^2$}

\affiliation{$^1$Leiden Observatory, Leiden University \\ Postbus 9513, 
NL-2300RA, Leiden, the Netherlands \\ email: {\tt haas@strw.leidenuniv.nl}\\
$^2$Sterrenkundig Instituut, University of Utrecht, \\ Princetonplein 5, 
NL-3584 CC Utrecht, the Netherlands \\ email: {\tt p.anders@uu.nl}}

\pubyear{2009}
\volume{266}  
\pagerange{XXX--YYY}
\jname{Star Clusters - Galactic Building Blocks throughout Time and Space}
\editors{Richard de Grijs \& Jacques Lepine, eds.}
\begin{document}

\maketitle

\begin{abstract}
If all stars form in clusters and both the stars and the clusters follow a 
power law distribution which favours the creation of low mass objects, 
then the numerous low mass clusters will be deficient in high mass stars. 
Therefore, the mass function of stars, integrated over the whole galaxy 
(the Integrated Galactic Initial Mass Function, IGIMF) will be steeper at 
the high mass end than the underlying IMF of the stars. We show how the 
steepness of the IGIMF depends on the sampling method and on the 
assumptions made for the star cluster mass function. We also investigate 
the O-star content, integrated photometry and chemical enrichment of 
galaxies that result from several IGIMFs, as compared to more standard IMFs.

\end{abstract}

\vspace{-1cm}

\section{Introduction}
In recent years, a series of papers 
(\cite{kroupaweidner03}, \cite{weidnerkroupa04}, \cite{weidnerkroupa05} and 
\cite{weidnerkroupa06}, WK06 from now on)
have proposed that the stellar content 
of an entire galaxy may not be well described by the same initial mass
function (IMF) that describes the distribution of stellar masses in the star 
clusters, where these stars form. The reason is that star clusters also form 
with a cluster mass function (CMF), which is a power law with a power law index of 
$\sim-2$. If the lower cluster
mass limit is very low (in the mass range that is also occupied by 
single stars, i.e.
below 100 \msun, and under the assumption that the most massive star in a 
cluster cannot be more massive
than its host cluster), the low mass clusters will be deficient in high 
mass stars. Therefore,
if the stellar content of all clusters is added up, resulting in the 
Integrated Galactic
Initial Mass Function (IGIMF), the distribution of stellar masses may be 
steeper at the
high mass end, depending on the exact shape of the CMF.

The exact form of the IGIMF could have profound implications for the integrated 
properties of galaxies. High mass stars are important for galaxies, both in terms of 
their energy output (massive stars are bright and explode in supernova explosions) 
and in terms of the chemical enrichment. Assuming an IGIMF instead of an IMF, the she supernova rate and chemical enrichment 
of galaxies 
are studied by \cite{goodwinpagel05}, the relation between H$\alpha$ and UV luminosity 
in star forming disks by \cite{pflammaltenburgkroupa08}. \cite{recchi09} investigates 
[$\alpha/$ Fe] abundance ratios for IGIMFs.

In most of these earlier studies, the assumed CMF and sampling method are kept the
same. For sampling they use the method `sorted sampling', which was put forward by
WK06, as they found it to best fit the relation between the most massive 
star in a cluster and the cluster mass. The CMF was assumed to be a pure power-law with index -2.2 and 
a lower cluster mass limit of 5\msun. 

In this contribution, we will vary the sampling method and the CMF parameters and 
investigate the resulting IGIMFs. Also, we will estimate the observability of the 
effects of sampling method and CMF variations through the number count of O-stars 
as will be observed by GAIA, the integrated photometry of distant galaxies and the 
chemical enrichment of galaxies (in terms of total gas phase metallicity).
 
\vspace{-.5cm}

\section{Constructing IGIMFs}
\subsection{Sampling method}
As a default choice for the CMF, we will use the same CMF as used before: 
\\${\rm d}N / {\rm d}M_{cl} \propto M_{cl}^{-2.2}$, with a lower mass limit of 5\msun.
For computational simplicity we use the Salpeter IMF for stars, between 0.1 and 100\msun.

We test several sampling methods (italic parts are the descriptions used in the figure 
later on):
\begin{enumerate}
\item Random sampling until the cluster mass is reached. As this will never get exactly
to the cluster mass aimed for, we either include the last star ({\it stop after}), 
exclude it ({\it stop before}), only exclude 
it if that brings the total mass in stars closer to the cluster mass aimed for than 
including it ({\it stop nearest}) or excluding the last star (which first surpassed 
the cluster mass) at a 50\% probability ({\it stop 50/50}).  
\item Sampling a specific number of stars, estimated from the total cluster mass and 
average mass in the IMF (with the maximum stellar mass limited to the cluster mass or not ({\it number} and 
{\it number unlimited})).
\item The `{\it sorted sampling}' method of WK06.
\item {\it Analytic sampling}, in which we assume that the IMF in a cluster is always
sampled analytically, excluding stochastic effects.
\end{enumerate}

For more detailed descriptions, see \cite{haasanders09}. We use a Monte Carlo approach in order to make the numerical scatter smaller than the difference between IGIMFs. We investigate the 
dependence of the results on the underlying IMF (we test for Kroupa) and the results 
are virtually indistinguishable.

\subsection{Cluster mass function}
As the main cause of the lack of high mass stars are the low mass clusters, the 
behaviour of the CMF at the low mass end is very important. The default choice is 
an extrapolation of the observed CMF at higher masses ($M_{cl} > \sim 10^3$\msun), down
to the mass of the smallest observed star forming clumps in the Taurus-Auriga region.
We therefore test, besides 
the default choice given above, also a lower limit of 1\msun\ and 50\msun\ for 
the clusters, and power law indices of -1.8, -3.2 and -4.2 (effectively changing the 
number of low mass clusters, compared to high mass clusters).


\section{Resulting IGIMFs}
\begin{figure}[h]
\centering
\begin{tabular}{ccc}
 \includegraphics[width=.45\linewidth]{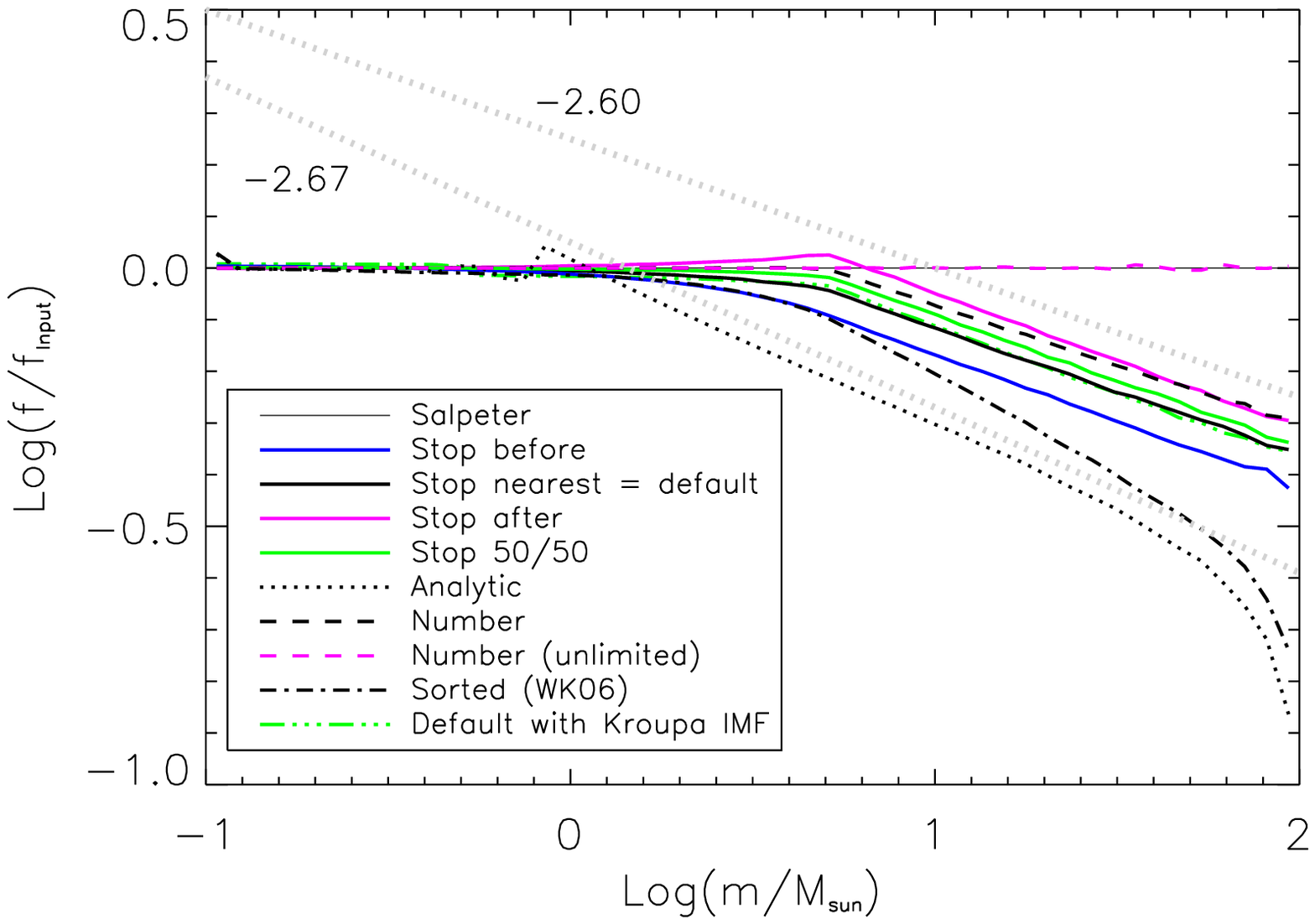} && 
 \includegraphics[width=.45\linewidth]{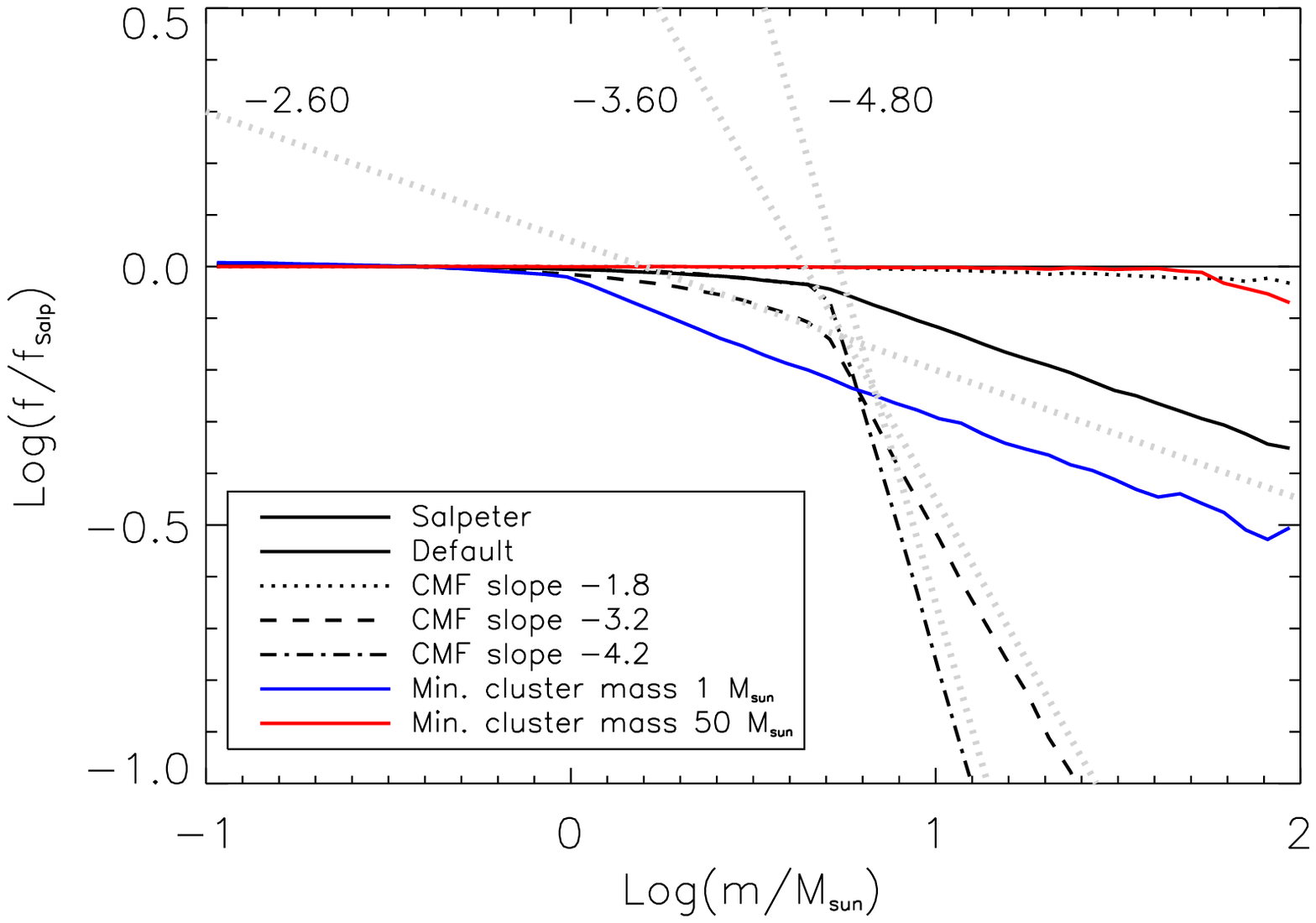}
\end{tabular}
 \caption{The resulting IGIMFs (normalized to the input IMF) 
	  for different sampling methods (left) and CMFs (right). 
          The indications correspond to the descriptions in the text. The dotted lines 
	  with indices indicate the steepness that line would have in an IGIMF.}
   \label{fig:igimfs}
\end{figure}

In Fig.~\ref{fig:igimfs} we show the resulting IGIMFs from different sampling methods
(left) and different CMFs (right). The value of the IGIMF is divided by the value for 
the underlying IMF, in order to enhance the visibility of the differences. Also, dotted lines indicate 
the slopes the different IGIMFs would have. 

It can be clearly seen that the different sampling methods give different IGIMFs.  
The recovered CMFs after filling the clusters with stars are indistinguishable 
from the input CMF. Depending on the method, the steep end has power law indices 
of about -2.6, as compared to the input -2.35. The `sorted sampling' and `analytic 
sampling' get even slightly steeper and deviate strongly from a power law shape.

From the Monte Carlo realisations using different CMFs, it can be seen that the 
slope of the CMF determines the high mass end slope of the IGIMF, whereas the minimum 
cluster mass sets the stellar mass at which the steepening sets in. If clusters do not 
typically go down in mass all the way to the mass range of stars, sampling effects in the 
IGIMF will be negligible. The low mass end of the CMF is badly constrained. The IGIMF could be obtained observationally, giving an indirect measure of the low mass end of the CMF (under the assumption of a sampling method and an underlying IMF).

\vspace{-0.5cm}

\section{Galaxy properties from IGIMFs}
Using the IGIMFs as IMFs in galaxy evolution models we are able to investigate the integrated 
properties of galaxies, if their total stellar initial mass distribution is given by an IGIMF 
rather than a more standard IMF. The high mass end of the stellar mass distribution is 
important for galaxies, as their light in many pass bands is dominated by the most massive stars for a wide range of star formation histories, their 
chemical composition depends on the metals expelled by the massive stars and the ISM is 
strongly influenced by the energy output of the massive stars exploding as supernovae.

\subsection{The number of O-stars in the Galaxy}
As the most pronounced effect appears for the most massive stars, we first estimate the 
difference in massive star content of the galaxies. Specifically, we estimate the number of 
O-stars (with masses $> 17$\msun) that GAIA will be able to observe. Assuming that GAIA 
will observe 10\% of all O-stars in the Galaxy, that they live for 10 Myr and that in the 
past 10 Myr the average SFR of the Galaxy was 1 \msun/yr, GAIA will be able to 
rule out several of the very extreme IGIMFs (from e.g. steep CMFs and sorted sampling) with 
very high significance. Judging between the several different sampling methods is difficult, 
as the differences are of the order of $1\sigma$, for purely poissonian errors on the number 
counts.

If our assumptions regarding the observed fraction of O-stars, the SFR and or the lifetimes 
of O-stars are inaccurate, the resultant number of O-stars varies linearly with either of 
them. The errors on the numbers scale with the square root of the numbers. As the numbers 
we calculate are of the order of 1000 to 2000, and the differences an order of magnitude 
smaller, the combined effect of our assumptions may be off by a factor of a few and the 
differences between all IGIMFs will still be of more than $2\sigma$ significance.

\subsection{Integrated photometry of galaxies}
Using the {\sc galev} evolutionary synthesis models (\cite{galev1, galev2}) we are able to 
assess observable properties of galaxies, in which the stellar initial mass distribution is 
given by IGIMFs, instead of more popular IMFs. The {\sc galev} models follow the chemical 
enrichment and photometric evolution of galaxies self-consistently in closed-box models of 
galaxy evolution. They are able to match a wide range of observed properties of galaxies 
of all Hubble types, like stellar mass, star formation and multi-band photometry.

The ingredients of the models are star formation histories which depend on the gas mass 
present (no in- and outflows are assumed, nor are they necessary) and Hubble Type, 
spectral synthesis, and yield tables. Subsequent generations of stars are formed from 
the gas enriched by earlier generations.

For all Hubble types, the difference in integrated photometry in any band is smaller than 
the observed galaxy-to-galaxy scatter as present in the HyperLeda 
database\footnote{http://leda.univ-lyon1.fr/} (\cite{hyperleda}). Therefore, integrated 
photometry of galaxies will not be a discriminant between IMFs and/or IGIMFs.

\subsection{Chemical enrichment}
Many heavy elements are mainly produced in supernova explosions of Type II (including Ib,c) 
and in the winds of massive AGB stars. With a deficiency in massive stars, as expected from 
the IGIMF, less production 
of metals is to be expected. The {\sc galev} models do not follow the chemical enrichment on 
an element by element basis, but rather follow the total metallicity. Element abundances can 
then be obtained by assuming solar abundance ratios.

With IGIMFs as determined above, metallicities of galaxies vary by half a dex at given gas 
mass fraction of the galaxies. For the IGIMFs resulting from extremely steep CMFs, the 
difference can be up to 2 orders of magnitude. At given gas mass fractions, the HyperLeda 
database gives a scatter of about half a dex as well. Therefore, the most extreme CMFs can 
be ruled out by the observed gas phase metallicities. In order to differentiate between 
several IGIMFs more precise measurements of metallicities are necessary. The difference 
between sampling methods is comparable to, or slightly larger than the spread that results 
from using individual metallicity measurements for the same galaxy, and is thus a possible 
discriminant between IGIMFs, provided that absolute determinations of gas phase metallicities 
are reliable.


\end{document}